\documentclass[]{raa_twocolumn}     
\usepackage{graphicx,times}
\usepackage{natbib}
\usepackage{amssymb,amsmath}
\bibpunct{(}{)}{;}{a}{}{,}

\def\mc{{{\mathcal{M}}}_{\rm c}}
\newcommand{\appropto}{\mathrel{\vcenter{
  \offinterlineskip\halign{\hfil$##$\cr
    \propto\cr\noalign{\kern2pt}\sim\cr\noalign{\kern-2pt}}}}}

\usepackage{multirow}
\usepackage{cancel}
\graphicspath{{figs/}}

\usepackage[a4paper=true,pagebackref=true]{hyperref}
\hypersetup{pdftitle = The title of my PDF, pdfauthor = My name, pdfsubject= The subject, pdfkeywords = keyword1 keyword2 keyword3} 
\hypersetup{colorlinks = true, linkcolor = green, anchorcolor = red, citecolor = blue, filecolor = red, pagecolor = red, urlcolor = red}

\begin{document}

   \title{On Dark Gravitational Wave Standard Sirens as Cosmological Inference and Forecasting the Constraint on Hubble Constant using Binary Black Holes Detected by Deci-hertz Observatory}

   \setcounter{page}{1}

   \author{Ju Chen\inst{1,2}, Changshuo Yan$^\ast$\inst{1,2}, Youjun Lu$^\dagger$\inst{1,2}, Yuetong Zhao\inst{1,2}, Junqiang Ge\inst{1}  }

   \institute{ National Astronomical Observatories, Chinese Academy of Sciences, 20A Datun Road, Beijing 100101, China; {\it $^\ast$\,yancs@nao.cas.cn; $^\dagger$\,luyj@nao.cas.cn}\\
        \and
             University of Chinese Academy of Sciences, 19A Yuquan Road, Beijing 100049, China\\
\vs \no
}

\abstract{Gravitational wave (GW) signals from compact binary coalescences can be used as standard sirens to constrain cosmological parameters if their redshift can be measured independently by electromagnetic signals. However, mergers of stellar binary black holes (BBHs) may not have electromagnetic counterparts and thus have no direct redshift measurements. These dark sirens may be still used to statistically constrain cosmological parameters by combining their GW measured luminosity distances and localization with deep redshift surveys of galaxies around it. We investigate this dark siren method to constrain cosmological parameters in detail by using mock BBH and galaxy samples. We find that the Hubble constant can be constrained well with an accuracy $\lesssim 1\%$ with a few tens or more of BBH mergers at redshift up to $1$ if GW observations can provide accurate estimates of their luminosity distance (with relative error of $\lesssim 0.01$) and localization ($\lesssim 0.1~\mathrm{deg}^2$), though the constraint may be significantly biased if the luminosity distance and localization errors are larger. 
We also introduce a simple method to correct this bias and find it is valid when the luminosity distance and localization errors are modestly large. 
We further generate mock BBH samples, according to current constraints on BBH merger rate and the distributions of BBH properties, and find that the Deci-hertz Observatory (DO) in a half year observation period may detect about one hundred BBHs with signal-to-noise ratio $\varrho \gtrsim 30$, relative luminosity distance error $\lesssim 0.02$, and localization error $\lesssim 0.01~\mathrm{deg}^2$. By applying the dark standard siren method, we find that the Hubble constant can be constrained to the $\sim 0.1-1\%$ level using these DO BBHs, an accuracy comparable to the constraints obtained by using electromagnetic observations in the near future, thus it may provide insight into the Hubble tension. We also demonstrate that the constraint on the Hubble constant applying this dark siren method is robust and does not depend on the choice of the prior for the properties of BBH host galaxies.
}

   \authorrunning{J. Chen et al. }            
   \titlerunning{Dark Gravitational Wave Standard Sirens}  
   \maketitle

%
\section{Introduction}
\label{sec:intro}

The Hubble constant ($H_0$), which measures the current expansion rate of the Universe, plays a fundamental role in state of the art cosmological models. An accurate measurement of $H_0$ is crucial for understanding the Universe, and numerous methods have been developed to achieve this goal \citep{Baade1944, Blandford1992, Riess1998, Perlmutter1999, Freedman2001, Spergel2007, Freedman2010, Beutler2011, Jackson2015}. Most notably, with the latest Planck cosmic microwave background (CMB) data, $H_0$ is determined to be $67.4\pm0.5 {\rm km\,s}^{-1} {\rm Mpc}^{-1}$ with a precision of $\sim 1\%$, assuming a standard $\Lambda$CDM cosmology \citep{PlanckCollaboration2018a}, while it is measured to be $74.03 \pm 1.42 {\rm km\,s}^{-1} {\rm Mpc}^{-1}$ by relying on the distance ladder measurement with a precision of $\sim 2\%$ \citep{Riess2018, Riess2019}. These estimates are discrepant by more than $4\sigma$. The result measured with lensed quasars recently is also different from the Planck measurement by $3\sigma$, and together with distance ladder, it suggests a $5\sigma$ tension between $H_0$ measurements obtained by early- and late-Universe probes \citep{Wong2020}. This tension may be an indicator of unknown systematic bias in current methods or new physics beyond our current standard cosmological models. It is therefore extremely important to find new and independent ways to measure the Hubble constant.

Compact binary coalesces may be used as ``standard sirens'' to measure $H_0$ and other cosmological parameters  \citep{Schutz1986}, if their redshift (information) can be obtained by independent method(s), as the luminosity distances of these sources can be directly measured from their gravitational wave (GW) signals. This method has been intensively investigated in the literature by using mock data \citep{Holz2005, Nissanke2010, Zhao2011, Taylor2012, Nissanke2013, Tamanini2016a, Liao2017, Chen2018, Li2019, Zhao2020, You2021, Wang2021}. The advantage of this method has been demonstrated by using observations of the first binary neutron star (BNS) merger GW170817, the redshift of which was measured via the discovery of its electromagnetic (EM) counterpart and host galaxy (NGC 4993) \citep{Abbott2017, Hotokezaka2019}.

To apply the ``standard siren'' method, it is necessary to have redshift measurements of binary coalescences by utilizing their EM counterparts since they cannot be obtained directly from the GW observations due to the degeneracy between chirp mass and redshift. However, it is not an easy task to identify their EM counterparts and/or host galaxies due to large uncertainties in the sky localization given by the GW signals and the small opening angle/faintness of the EM emission. Moreover, a large fraction of the GW sources detected by ground-based detectors would be mergers of binary black holes (BBHs), which are even not expected to be accompanied by bright EM counterparts.

The planned third generation ground-based GW detectors (e.g., Einstein Telescope: ET, and Cosmic Explorer: CE) and future space GW detectors (e.g., DECIGO and Deci-hertz Observatory, DO) are expected to detect a large number of compact binary mergers with high signal-to-noise ratio (S/N) at redshift up to $10$, and will yield $\sim 10^4\sim10^8$ detections of BBH events per year \citep{Abernathy2011, Abbott2017, ArcaSedda2020}. Most of these mergers, if not all, will not have detectable EM counterparts and thus redshift information. It is then of great interests to investigate whether these dark ``standard sirens'' can be used to probe cosmology and constrain cosmological parameters.

\citet{Schutz1986} originally considered that each galaxy (cluster) within the localization error volume can be taken as a potential host candidate, and by combining different GW events, only one galaxy (cluster) in each error volume will give consistent constraints on cosmological parameters. The effectiveness of this method has been discussed recently in various cases including both ground-based \citep{DelPozzo2012, Chen2018, Gray2020} and space-based \citep{MacLeod2008, Petiteau2011, DelPozzo2018, Wang2020} detectors. \citet{Yu2020} further investigated the capabilities of identifying host galaxy groups instead of host galaxies in considering the hardness of carrying out deep galaxy survey. Such a method is also applied to the real observational data as presented in \citet{Fishbach2019} for GW170817 without considering the EM counterpart, and \citet{TheDESCollaboration2019} for GW 170814.  \citet{TheLIGOScientificCollaboration2021} also presented the result obtained by combining observations of the current standard siren GW170817 and all other detected dark GW events.

In this paper, we investigate the validity and limitation of relying on dark ``standard sirens'' (mainly mergers of stellar BBHs in this paper) to constrain cosmological parameters and quantify the accuracy of such a method and the selection criteria  for the sample of dark ``standard sirens'' to be used for cosmological application. The paper is organized as follows. In Section~\ref{sec:method}, we introduce the framework used in our analysis. In Section~\ref{sec:mockdata}, we describe the detailed method for generating GW sources and host galaxy mock data based on a simple model for the evolution merger rate density and the Millennium database. In Sections~\ref{sec:result} we check the robustness of the method, and discuss the systemic bias we noticed in detail and also introduce a correction factor. In section~\ref{sec:DO}, we further investigate whether the DO could provide a large number of GW events with sufficiently smaller errors of luminosity distance measurements and localization and test the ability of DO observations to constrain the cosmological parameters via dark sirens. Conclusions are summarized in Sections~\ref{sec:conclusion}

Throughout the paper, we assume a flat $\Lambda$CDM universe, with the Hubble constant $H_0$, matter fraction of $\Omega_{\rm m}$, and cosmological constant $(1-\Omega_{\rm m}$). We adopt the conventional dimensionless parameter $h = H_0/100~{\rm km~s}^{-1}\,{\rm Mpc}^{-1}$ to denote the Hubble constant.

\section{The general framework}
\label{sec:method}

In this section, we briefly introduce the general framework to infer cosmological parameters by utilizing the dark standard sirens, e.g., mergers of BBHs without EM counterparts and even mergers of faraway BNSs with too faint EM counterparts to be detected. GW signals provide estimates of the luminosity distances and sky localization of these dark sources but not the redshift. EM galaxy surveys can provide host candidates for these GW sources in the cosmic volume defined by the errors of luminosity distance and sky location of each source. The combination of the GW measurements and galaxy surveys may enable constraints on cosmological parameters by using these dark standard sirens \citep[][see also \citealt{DelPozzo2012, Petiteau2011}]{DelPozzo2018}. 

Consider a set of $n$ GW observations ${\boldsymbol D} = ({\boldsymbol D}_1, \cdots, {\boldsymbol D}_i, \cdots, {\boldsymbol D}_n)$ and an all sky galaxy catalog ${\boldsymbol g}$. For each GW observation ${\boldsymbol D}_i$, we have all the parameters characterizing the binary system obtained from the GW signals except that the chirp mass and redshift are degenerate. We assume that each GW source is hosted in a galaxy and the redshift of a GW event can be given approximately by its host galaxy redshift, by ignoring the motion of the GW source itself with respect to its host galaxy.\footnote{This relative motion may be typical of the rotation velocity or velocity dispersion of the host galaxy, say, $\sim 200~{\rm km}~s^{-1}$ for an $L_\ast$ galaxy, smaller than the galaxy's peculiar velocity, which can be safely ignored.} In the case of no EM counterpart identification, the redshift (probability distribution) may be still inferred from galaxies located in the sky area of the GW source with properties (e.g., R.A. $\varphi_i$, decl. $\theta_i$, redshift $z_i$, stellar mass $M_\ast$, etc.) determined by deep galaxy surveys. We assign a probability for each galaxy in the localization error volume (defined by the localization error $\Delta\Omega$ and luminosity distance $\delta d_{\rm L}$) of the GW source so that the probability distribution of the GW source redshift can be obtained. Note here that $\delta d_{\rm L}$ and $\Delta \Omega$ given by GW observations usually adopt the errors associated with $90\%$ confidence level. If $\Delta \Omega$ and $\delta d_{\rm L}$ calculated from GW signals are sufficiently small, the number of possible galaxies in the error volume ($\propto  \Delta \Omega d^2_{\rm L} \delta d_{\rm L}$) would be small and thus the inferred redshift (distribution) can be quite close to the real one.

For simplicity, we assume the sky localization area for a GW event is circular in the following analysis although the one given by GW observation is normally elongated and even irregular. Considering that our results are only affected by the distribution of galaxies along the redshift direction but not the direction transverse to the line of sight (LOS), the approximation of a circular localization area should not have any significant effect on our results, though one  must consider this shape in practical studies.

With the information from both GW observations and galaxy surveys described above, the posterior probability distributions for the cosmological parameters, including the Hubble constant ($H_0$ or $h$), fraction of matter content ($\Omega_{\rm m}$) and others, may be estimated by
\begin{equation}
\label{eq:bayes}
p(H|{\boldsymbol D},I) = \frac{p(H|I)p({\boldsymbol D}|H,I)}{p({\boldsymbol D}|I)},
\end{equation}
according to the Bayes' theorem. Here evidence ${\boldsymbol D} = ({\boldsymbol D}_1,\cdots,{\boldsymbol D}_i,\cdots,{\boldsymbol D}_n)$ with ${\boldsymbol D}_i$ representing the observational data, including the dataset ${\boldsymbol d}_i$ given by GW observations and the redshift ${\boldsymbol z}_i =(z_{i,1},\cdots,z_{i,j},\cdots)$ of host galaxies in the localization error volume of each GW source $i$, $I$ signifies all the additional information and assumptions, $p(H|I)$ is the prior probability distribution for cosmological parameters, $p({\boldsymbol D}|H,I)$ is the likelihood function for evidence ${\boldsymbol D}$, $p({\boldsymbol D}|I)$ is the probability distribution for evidence ${\boldsymbol D}$ under all possible parameter spaces and can be treated as a normalization constant here.
Since GW events should be independent of each other, the likelihood above can be rewritten as the product of the likelihoods for individual events ${\boldsymbol D}_i$, i.e.,
\begin{equation}
p({\boldsymbol D}|H,I) = \prod_{i=1}^n p({\boldsymbol D}_i|H,I).
\end{equation}

The likelihood for a single GW event is obtained by marginalization over all different parameters $\mathbf{x}$ of the source system, i.e.,
\begin{equation}
p({\boldsymbol D}_i|H,I) = \int d \mathbf{x}\, p({\boldsymbol D}_i|\mathbf{x},H,I)p(\mathbf{x}|H,I).
\end{equation}
For the consideration in the present paper, these parameters include the celestial coordinates $\theta$, $\varphi$, luminosity distance $d_{\rm L}$, and redshift $z$, i.e., $\mathbf{ x}\equiv \{\theta,\varphi, d_{\rm L}, z\}$. We assume that the distribution of GW source luminosity distance $d_{\rm L}$ is independent of sky location $(\theta,\varphi)$, then we have
\begin{equation}
p({\boldsymbol D}_i|H,I) = \int d\mathbf{x} p({\boldsymbol D}_i|\mathbf{x},H,I) p(d_{\rm L}|z,H,I) p(\theta, \varphi, z|H,I).
\label{eq:indv_start}
\end{equation}

The likelihood for each GW event $p({\boldsymbol D}_i|\mathbf{x},H,I)$ is assumed to be Gaussian distribution centered on the luminosity distance $\bar{d}_{{\rm L}_i}$ with a standard deviation of $\sigma_{d_{{\rm L}i}}$ inferred from the GW signal, 
i.e.,
\begin{equation}
p({\boldsymbol D}_i|\mathbf{x},H,I) \propto \mathcal{N}(\bar{d}_{{\rm L}_i}, \sigma_{d_{{\rm L}_i}}) \propto \exp\left[- \frac{\left(d_{{\rm L}_i}(z,H) - \bar{d}_{{\rm L}_i}\right)^2}{2 \sigma^2_{d_{{\rm L}_i}}}\right].
\end{equation}
Here the luminosity distance of the host galaxies in the localization error volume of each GW source $i$ is expressed as
\begin{equation}
d_{{\rm L}_i}(z, H) = (1+z) \int_0^z \frac{cdz'}{H_0 E(z')},
\label{Eq:DLofz}
\end{equation}
with $E(z)=\sqrt{\Omega_{\rm m} (1+z)^3 + 1-\Omega_{\rm m}}$ for the assumed flat $\Lambda$CDM model. $\sigma_{d_{\rm L}}$ can be converted from $\delta d_{\rm L}$ given by the $90\%$ confidence level assuming a Gaussian distribution of the error. Therefore,
\begin{equation}
p(d_{\rm L}|z,H,I) = \delta(d_{\rm L}-d_{\rm L}(z,H)).
\end{equation}

With the assumption that GW sources are hosted in galaxies, the joint distribution of other parameters $p(\theta,\varphi, z|H,I)$ can be obtained by summing up the probability of all galaxies located in the localization error volume being the host. Here we simply choose a sum of $\delta$-functions over all individual galaxies, i.e.,
\begin{equation}
p(\theta,\varphi, z|H,I)= \sum_{j=1}^m \delta(z-z_j) \delta(\theta-\theta_j) \delta(\varphi-\varphi_j) p_j,
\label{eq:indv_end}
\end{equation}
where $j$ goes through all the cataloged galaxies located in the localization error volume of each GW event ${\boldsymbol D}_i$, $m$ is the total number of these galaxies and it may be significantly different for different GW events, $p_j$ is the weight assigned to each of these galaxies according to additional information, if available, about the probability distribution of host galaxy properties, and we assume it does not depend on $\theta_j$, $\varphi_j$, or $z_j$.
In principle, $p_j$ is not known as a prior. The simplest way to consider this probability is to assume that the BBH merger rate for each galaxy is proportional to its stellar mass $M_\ast$. In this case, the probability of a galaxy with $M_\ast$ to be the host is
\begin{equation}
p_j \propto \frac{M_{\ast,j}}{\int dM_\ast M_\ast \frac{dN}{dM_\ast}} \appropto \frac{M_{\ast,j}}{\sum_j M_{\ast,j}} \propto M_{\ast,j},
\end{equation}
where $dN/dM_\ast$ is the stellar mass function of galaxies, which in our case for all galaxies in the error volume of each GW event is $\sum_j \delta(M_\ast-M_{\ast,j})$.

Note here we omit the additional subscript $i$ for each GW event in the above equation and hereafter for simplicity, but we should keep in mind that for each GW event ${\boldsymbol D}_i$, $m$ can be quite different.

A more realistic distribution of host galaxy properties may be statistically inferred by future GW detection of a large number of BBH mergers. Such a distribution can also be derived by considering detailed formation and evolution of BBHs along with galaxies in the Universe \citep{Cao2018,  Artale2019}. In this paper, we simply use $p_j \propto M_{\ast,j}$ to assign the host galaxy for each GW event when generating the mock sample (for details see section~\ref{sec:mockdata}). For the inference of cosmological parameters, we also assume $p_j$ is known as a prior, and set the weight according to galaxy stellar mass if not otherwise stated. The effects of different choices of $p_j$ on the cosmological parameter inference will be further discussed in section~\ref{sec:discussion}.

According to equations~\eqref{eq:indv_start} to \eqref{eq:indv_end}, the likelihood for an individual GW event is then
\begin{align}
p({\boldsymbol D}_i|H,I)  = & \int d \mathbf{x} ~\mathcal{N}(\bar{d}_{{\rm L}_i}, \sigma_{d_{{\rm L}_i}})
\delta(d_{\rm L}-d_{\rm L}(z,H)) \nonumber \\
& \sum_{j=1}^m \delta(z-z_j)
\delta(\theta-\theta_j)\delta(\varphi-\varphi_j)p_j.
\end{align}
By integrating over parameters $d_{\rm L}$, $\theta_j$ and $\phi_j$, we have
\begin{equation}
p({\boldsymbol D}_i|H,I) \propto \sum_{j=1}^m \exp \left[- \frac{\left(d_{{\rm L}_i}(z_j, H) - \bar{d}_{{\rm L}_i}\right)^2}{2\sigma^2_{d_{{\rm L}_i}}}\right] p_j,
\label{eq:likelihood_noprop}
\end{equation}

In the above Equations~\eqref{eq:indv_end} and \eqref{eq:likelihood_noprop}, the uncertainty in the redshift measurement for each galaxy $z_j$, e.g., due to peculiar velocity, is not considered. However the influence of peculiar velocities is not negligible for nearby galaxies. To account for the effect of peculiar velocity, we assume the uncertainty of redshift introduced by peculiar velocity is a normal distribution around the true redshift, then we have
\begin{align}
p(\theta,\varphi, z|H,I)= \sum_{j=1}^m \exp\left[ -\frac{\left({z_j}_{\rm obs} - z\right)^2}{2 \sigma^2_{z_{\rm pv}}}\right] 
 \delta(\theta-\theta_j) \delta(\varphi-\varphi_j) p_j, \nonumber \\
\end{align}
where ${z_j}_{,\mathrm{obs}}$ is the observed redshift including peculiar velocity and can be obtained from the mock catalog \citep{Henriques2012}. The standard deviation $\sigma_{z_{\rm pv}}$ is set to $0.0015$ for low redshift galaxies ($z<0.1$) and $0.0015+0.0005z$ for galaxies with higher redshift. These values are chosen based on the distribution of redshift measurement error for all galaxies in the mocked galaxy catalog in \citet{Henriques2012}.

With the above assumption about redshift distribution of galaxies, the likelihood for a GW event becomes
\begin{align}
p(\boldsymbol{D}_i|H,I)   \propto   \int_{z^{\min}}^{z^{\max}} & dz \exp\left[-\frac{\left(d_{\rm L}(z, H) - \bar{d}_{L_i}\right)^2}{2 \sigma^2_{d_{L_i}}}\right] \nonumber \\
& \times \sum_{j=1}^m \exp\left[ -\frac{\left({{z_j}_{\rm abs} - z}\right)^2}{2 \sigma^2_{z_{\rm pv}}}\right] p_j. 
\label{eq:likelihood-prop}
\end{align}
The upper and lower limits ($z^{\min},z^{\max}$) of the integration are estimated as described in section~\ref{sec:errorbox} for the selection of host galaxy candidates.

To investigate the prospects of the dark siren method described above for future GW observations, below we generate mock samples of GW events with luminosity distance and localization measurements. Correspondingly, for each GW event we can get a list of galaxies as host candidates from the mocked galaxy catalog given by \citep{Henriques2012}. Then, we adopt the above framework for Bayes's inference to reconstruct cosmological parameters from the mock data. We adopt the nested sampling algorithm provided by {\tt dynesty} \citep{Speagle2020} to carry out the posterior distribution sampling. We first investigate the dark siren method by using mock data with a set of assumed uncertainties for the luminosity distance measurements and localization via GW observations, and then test with mock data simulated for future DO observations \citep{ArcaSedda2020}.

To generate mock catalogs of GW events, we assume a flat $\Lambda$CDM Universe with $h=0.73$, $\Omega_{\rm m}=0.25$. The parameters are chosen to be the same as the Millennium Simulation for consistency as the host galaxies are selected by using the galaxy catalogs obtained from the Millennium Simulation.

\section{Mock data for GW events and their host galaxies}
\label{sec:mockdata}

\subsection{Stellar binary black holes}
\label{sec:bbhsample}

We first generate mock samples of stellar binary black holes (BBHs) at $z<3$ by adopting the BBH merger rate density (and its evolution) inferred from the LIGO/Virgo observations after O3a \citep{ligoO3a}, i.e.,
\begin{equation}
R(z) = R_0(1+z)^\kappa,
\label{eq:Rz}
\end{equation}
where $R_0$ is the local merger rate density. According to the power law plus a peak model in \citet{ligoO3a}, we have $R_0 = 19.1 {\rm Gpc^{-3}yr^{-1}}$ and $\kappa = 1.3$. We ignore the difference of the cosmological model adopted for the estimation of $R_0$ from the ``true'' cosmological model (Millennium Simulation) considered in this paper for mocking GW events and their host galaxies, as current estimation on $R_0$ still has a large uncertainty. We assume that the BBH merger rate density follows equation~\eqref{eq:Rz} up to $z=3$ although it is obtained by only using LIGO/Virgo detected BBHs at low redshift.

\subsection{Host galaxies of GW events}
\label{sec:source}

In this subsection, we describe how to generate mock samples of GW events and their host galaxies. According to equation~\eqref{eq:Rz}, we can randomly generate a given number of BBH mergers with assigned redshift $z'$. We adopt the virtual observed light cones of \citet{Henriques2012}, constructed from the semi-analytic galaxy formation model \citep{Guo2011} based on the Millennium Simulation \citep{Springel2005}. In these light cones, galaxies with different stellar masses (or other properties) may not have the same probability to be the host galaxy of a GW event. In order to account for the dependence of a GW event on galaxy stellar mass, as discussed in section \ref{sec:method}, we generated a catalog of GW host galaxies from the simulated one given by \citet{Henriques2012}, by convolution of the probability of a galaxy with mass $M_\ast$ to be the GW host $p(M_\ast)$. By default, we will choose the distribution simply as $p(M_\ast) \propto M_\ast$,  and we will further discuss the influence of $p(M_\ast)$ in section \ref{sec:discussion}. For each BBH merger at $z'$, we may then find a galaxy in the new catalog for GW hosts with redshift $z$ that is the closest one to the BBH merger at $z'$. We replace the redshift of the GW event $z'$ by $z$, and its host has a mass of $M_\ast$. By doing so, we can generate catalogs of GW events with known host galaxies and perform further analysis. According to the light cones (assuming periodical boundary condition), we can get the sky location, stellar mass, real redshift $z$, observed redshift $z_{\rm obs}$ (including the effect of peculiar velocity), and apparent magnitude of each galaxy for each GW event and its host galaxy. Note here that galaxies with $M_\ast<10^9 M_{\odot}$ are excluded from our consideration, since they are rare and may be biased due to the resolution limitation of the Millennium Run.

As a test of the method, we first generate some mock samples (both at redshift $z<1$ and $z<0.1$) of GW events with assumed fixed uncertainties of the luminosity distance measurements and localization in section \ref{sec:result}. After that, we investigate the prospective of future observation with mocked detection of DO in section \ref{sec:DO}. In these cases, we first obtain the luminosity distance $d'_{\rm L}$ from the true redshift of the host galaxy of each GW event under the cosmological model ($h=0.73$, $\Omega_{\rm m}=0.25$). We then get the observed luminosity distance $\bar{d}_{\rm L}$ for these GW events by adding a random error to $d'_{\rm L}$, and the errors are chosen to be Gaussian distribution with standard deviation $\sigma_{d_{\rm L}}$ given by mock observations.

\subsection{Selection of host-galaxy Candidates for GW events}
\label{sec:errorbox}

Given the luminosity distance and its uncertainty $\bar{d}_{\rm L} \pm \delta d_{\rm L}$ of a GW event, we can get the corresponding redshift range for any given set of cosmology parameters. If the cosmology parameters are not known prior, the error range for the redshift of the GW event would be significantly larger as one must account for the uncertainties in the cosmological parameters.

We choose the prior for $h$ and $\Omega_{\rm m}$ as flat distributions in $[0.6, 0.8]$ and $[0.01, 0.5]$, respectively. With a given set of $(h,\Omega_{\rm m})$, according to the flat $\Lambda$CDM model, the luminosity distance and its error $\bar{d}_{\rm L} \pm \sigma_{d_{\rm L}}$ of GW event can be converted into a redshift range $[z^-,z^+]$. In this paper, we adopt a 3$\sigma_{d_{\rm L}}$ range to determine the redshift error range. Under the assumed priors for $h$ and $\Omega$, the minimum value of $z^-$ is from the model with $(h,\Omega_{\rm m})=(0.6,0.01)$, and the maximum value of $z^+$ is from the model with $(h,\Omega_{\rm m})=(0.8,0.5)$. Furthermore, galaxy surveys normally give the observed redshift of each galaxy, which deviates from the true one because of the contamination by peculiar velocity. To account for this contamination, the redshift error range is enlarged from $(z^-,z^+)$ to $(z^{\rm min},z^{\max})$, which includes the errors induced by the peculiar velocity as $z^{\min} = z^{-} - 3\sigma_{z_{\rm pv}}$, $z^{\max}=z^{-}+3\sigma_{z_{\rm pv}}$, with $\sigma_{z_{\rm pv}}$ also set to 0.0015 for low redshift galaxies ($z < 0.1$) and $0.0015+0.0005z$ for galaxies with higher redshift.

With the redshift range determined above (without fixing the cosmological parameters $h$ and $\Omega_{\rm m}$), the error volume  can be obtained by considering both the localization error  $\Delta \Omega$ and redshift range $[z^{\min}, z^{\max}]$. All galaxies falling within the error volume are then selected as potential hosts.

For the galaxy catalogs given by \citet{Henriques2012}, we use the all-sky galaxy map for GW events at low redshift ($z<0.15$), while the pencil-beam light cones assume a periodic expansion for events at high redshift. We also set a threshold on the $i$-band apparent magnitude limit of host galaxies as $25$ at high redshift.

\section{Robustness check of the method}

\label{sec:result}

\subsection{Systematic bias in the method}

We first investigate the dependence of cosmological constraints on the measurement errors of luminosity distance and localization of GW events. 
Note, for real observation the errors are normally smaller for nearby events and larger for more distant events. Here to demonstrate the dependence of the method on the observation errors, we will first ignore this and simply assume that all GW events have the same uncertainties in luminosity distances and localization. More realistic mock observation will be presented in section \ref{sec:DO} for DO.

\begin{figure*}
\centering
\includegraphics[width=0.47\textwidth]{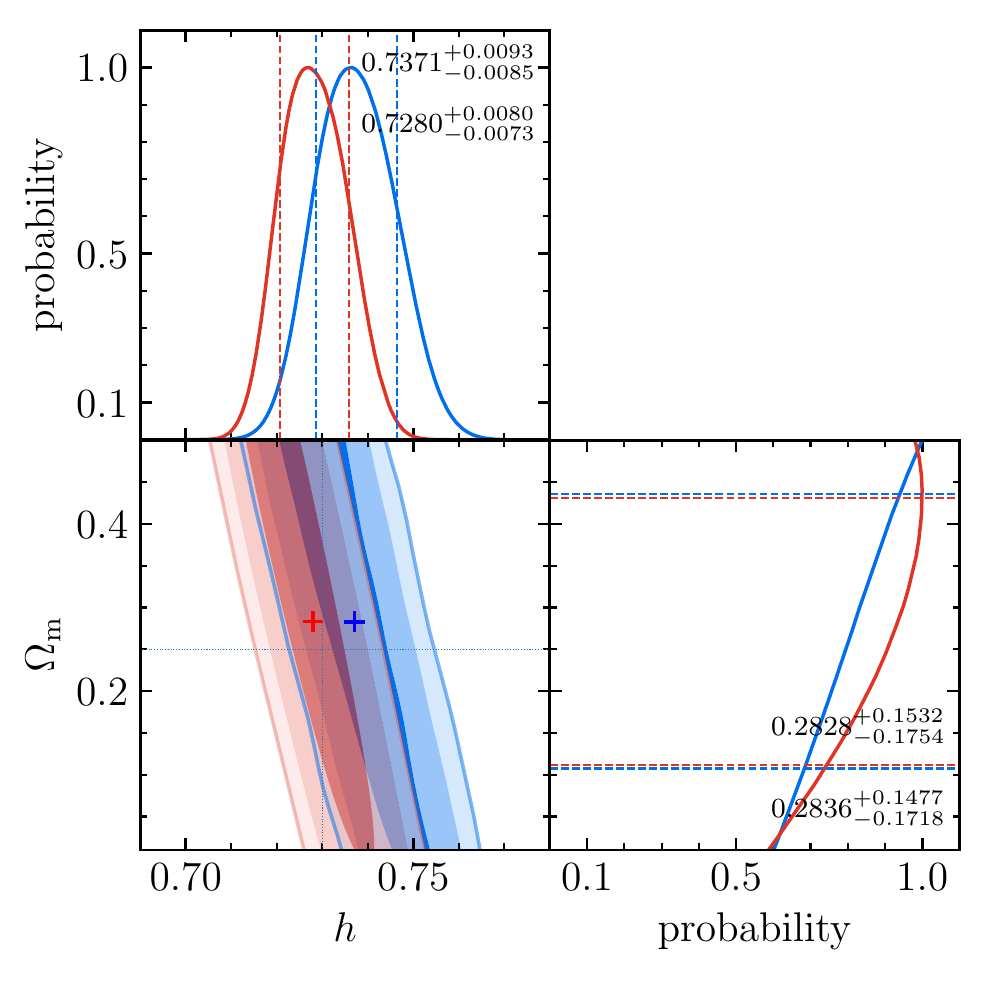}
\includegraphics[width=0.47\textwidth]{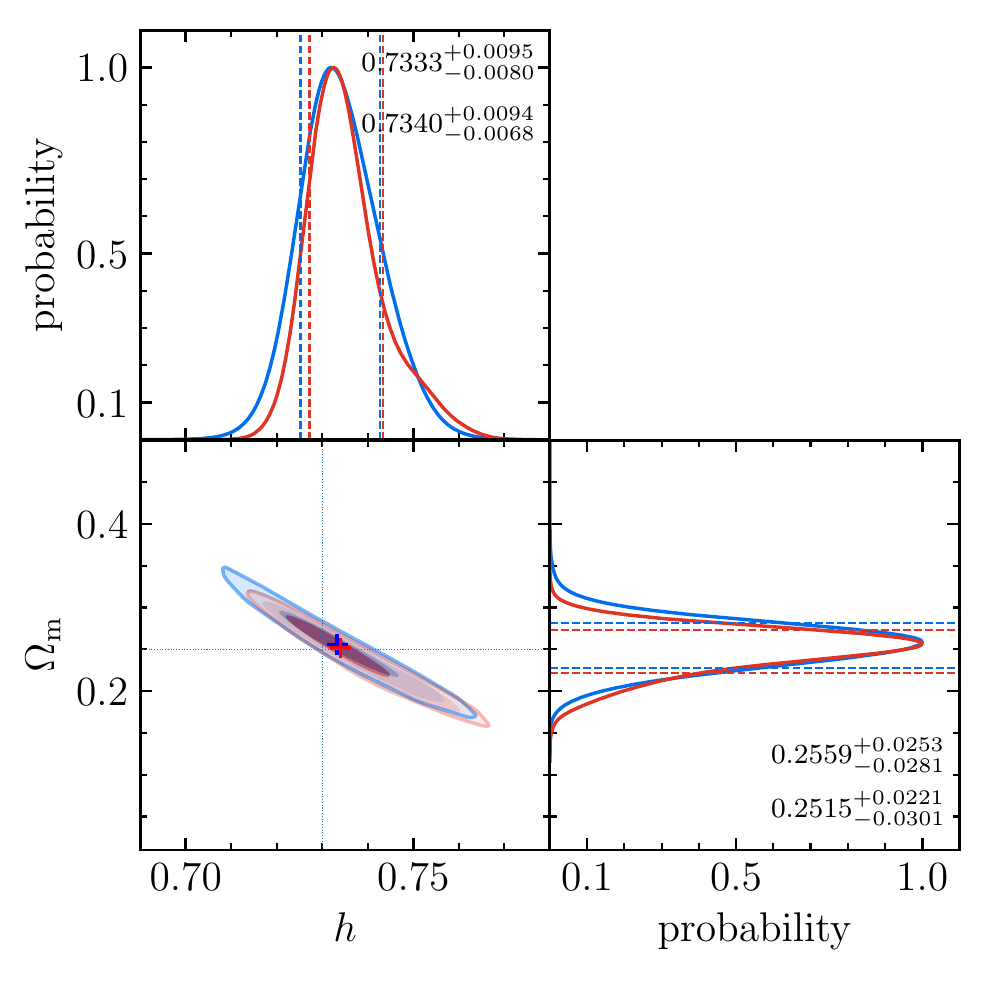}
\caption{
Posterior probability distributions of $h$ and $\Omega_{\rm m}$ obtained from mock GW events at redshift $z \leq 0.1$ (left panel) and $z \leq 1$ (right panel). The relative uncertainty for the luminosity distance $\delta {d_{\rm L}}/d_{\rm L}$ is set to $0.01$, and the sky localization uncertainty $\Delta\Omega$ is fixed at  {$5~\mathrm{deg}^2$} for mock GW events adopted at $z\leq 0.1$ and $0.1\mathrm{deg}^2$ at $z\leq 1$, respectively.
The blue and red contours in the bottom-left sub-figure of each panel signify the 1, 2 and 3$\sigma$ level constraints obtained from the method described in section~\ref{sec:method} without and with consideration of the correction factor introduced in following, respectively. The blue and red curves in the top and bottom-left sub-figures of each panel correspondingly depict the marginalized one-dimensional probability distributions of $h$ and $\Omega_{\rm m}$(in the horizontal direction), respectively.
Vertical and horizontal dashed lines indicate the input values of $h$ and $\Omega_{\rm m}$, and the median values of recovered parameters are shown as '+' symbols.
This figure affirms that the cosmological parameters can be constrained well when the luminosity distances and localization of the GW sources are measured with high precision, and with the correction factor $\alpha(H)$ introduced in section~\ref{sec:result-bias}, the constraints are only slightly improved.
}
\label{fig:f1}
\end{figure*}

\begin{figure*}
\centering
\includegraphics[width=0.47\textwidth]{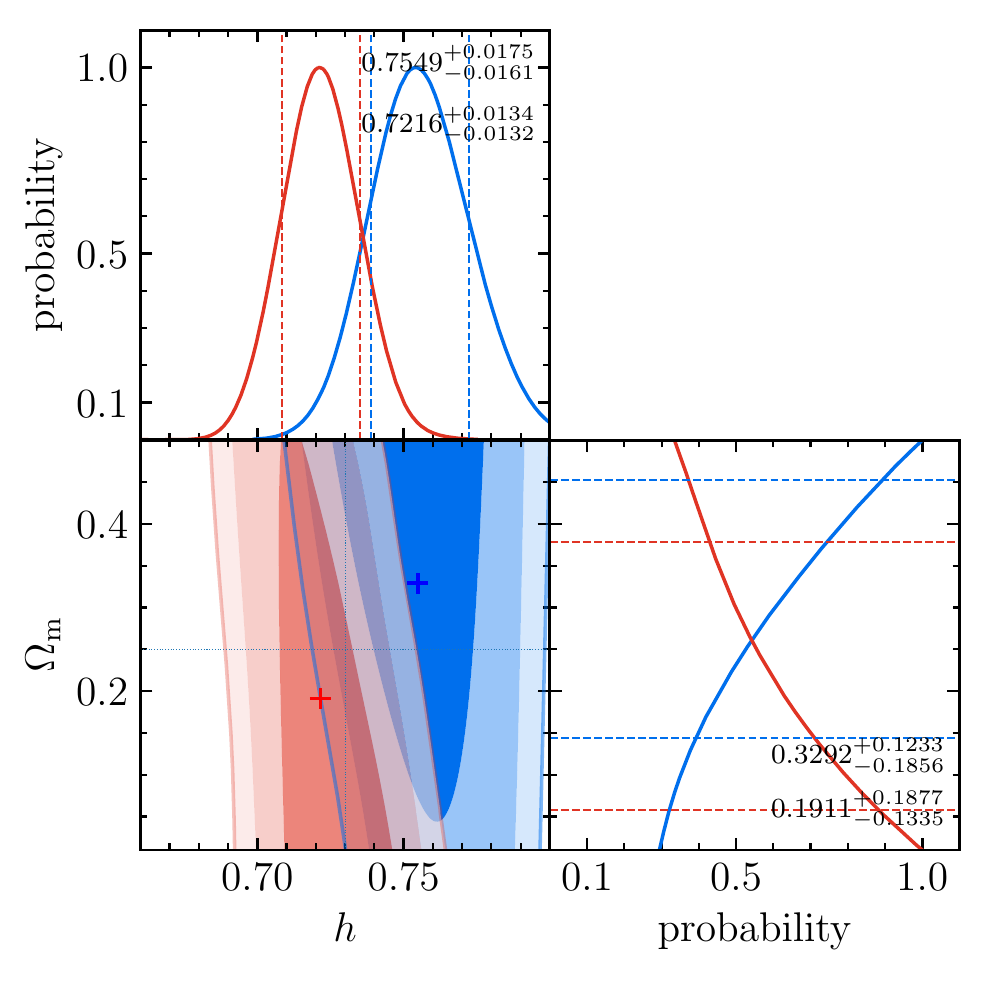}
\includegraphics[width=0.47\textwidth]{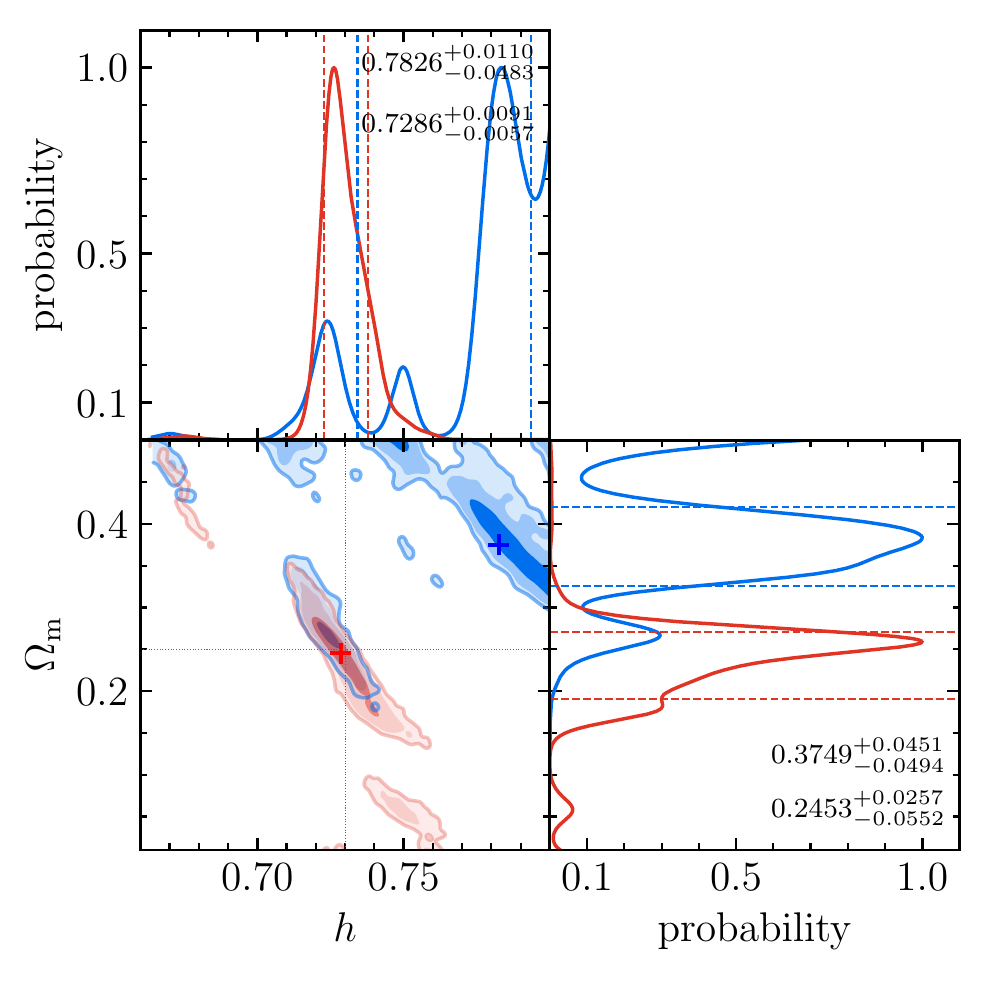}
\caption{
Legend is similar to fig.~\ref{fig:f1}, except that the measurement errors of the luminosity distances and localization areas for GW sources are set to $(\delta d_{\rm L}/d_{\rm L},\Delta \Omega)=(0.1,~5~\mathrm{deg}^2)$, and $(0.01,~1~\mathrm{deg}^2)$, respectively. 
This figure demonstrates that the inferred cosmological parameters are highly biased when the measurements of the luminosity distances and localization of GW sources are not precise enough (cf. Fig.~\ref{fig:f1}), and the constraints can be improved significantly once the bias correction discussed in section~\ref{sec:result-bias} is considered.
}
\label{fig:f2}
\end{figure*}

Figure~\ref{fig:f1} displays the posterior probability distributions of cosmological parameters $h$ and $\Omega_{\rm m}$ obtained by using $50$ mock GW events at redshift $z \leq 0.1$ (left panel) and $z\leq1$ (right panel). The uncertainties ($90\%$ level) of the luminosity distance measurements from the GW signals for all events are set as $\delta d_{\rm L}/d_{\rm L}=0.01$, while the sky localization uncertainties ($90\%$ level) are set as $\Delta\Omega = 5~\mathrm{deg}^2$ and $0.1~\mathrm{deg}^2$ for GW events applied in the left and right  panels, respectively. As shown with the blue color results, $h$ can be constrained with a precision of $\sim 1\%$ even if only $50$ dark sirens with redshift up to $1$ are adopted. However, $\Omega_{\rm m}$ cannot be constrained well, especially if only low redshift ($z<0.1$) GW events are available, which is consistent with \citet{DelPozzo2018}. Adopting $50$ GW events with redshift up to $1$, the constraint on $\Omega_{\rm m}$ can be improved but only reach a precision of $\sim 8\%$.

\begin{figure*} 
\centering
\includegraphics[width=0.49\textwidth]{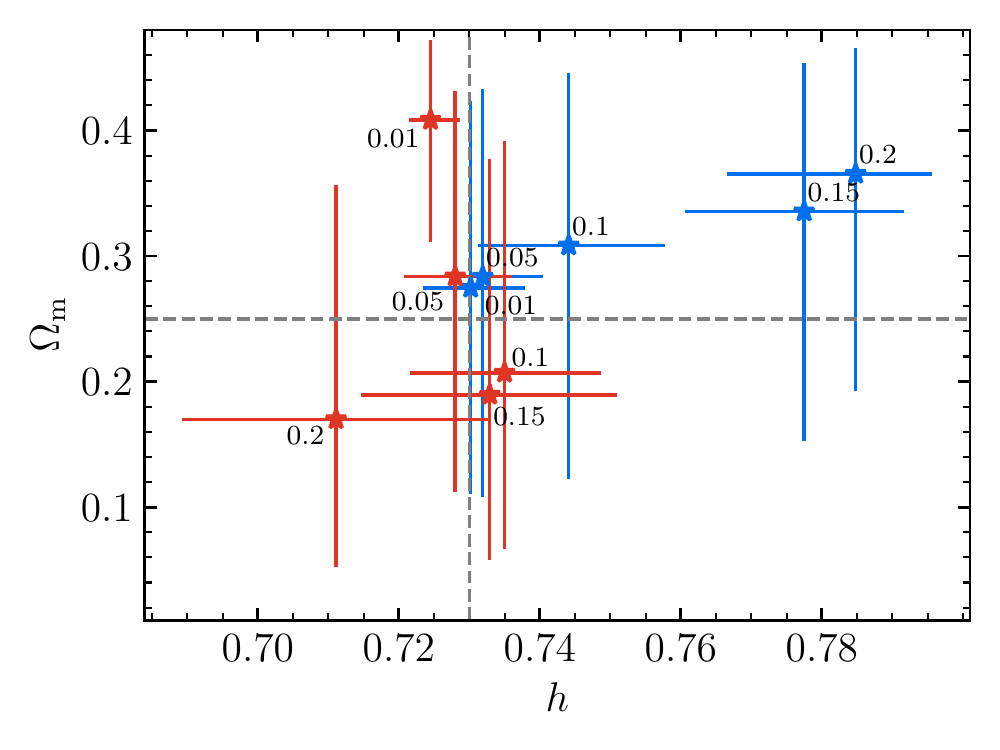}
\includegraphics[width=0.49\textwidth]{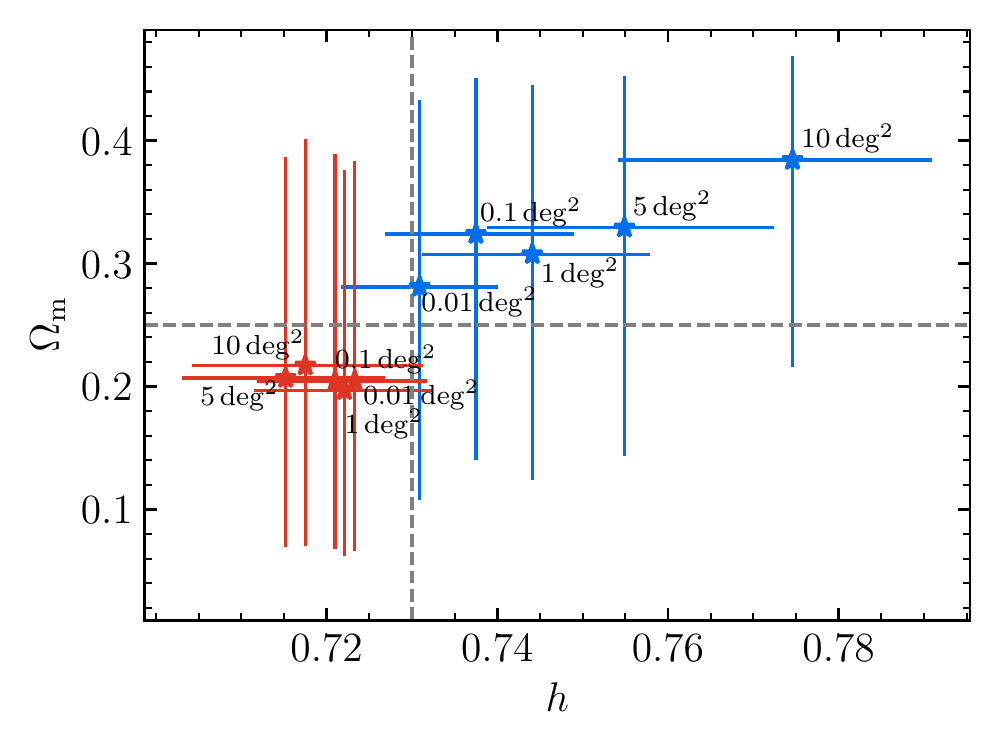}
\caption{
Recovered cosmological parameters (median value case of 5 runs) from $50$ mock GW events at $z<0.1$ by assuming different luminosity distance and localization measurement errors. In the left panel, the localization errors are all assumed to be $1\mathrm{deg}^2$, but the relative luminosity distance errors are assumed to be $\delta d_{\rm L}/d_{\rm, L}=0.01$, $0.05$, $0.1$, $0.15$ and $0.2$ as indicated, respectively. In the right panel, the relative luminosity distance errors are all assumed to be $0.01$, while the localization errors are assumed to be $0.01~\mathrm{deg}^2$, $0.1~\mathrm{deg}^2$, $0.1~\mathrm{deg}^2$, $1~\mathrm{deg}^2$, $5~\mathrm{deg}^2$ and $10~\mathrm{deg}^2$, as indicated, respectively. The error bars associated with each symbol mark the 1$\sigma$ error range of the constraint.
Blue and red stars indicate the results obtained without and with consideration of the bias correction (see section~\ref{sec:result-bias}), respectively. This figure suggests that the constraints can be improved significantly once the bias correction is considered for those samples with relatively modestly large luminosity distance and localization measurement errors.
}
\label{fig:f3}
\end{figure*}

Figure~\ref{fig:f2} displays the constraints obtained from $50$ mock GW events with $(\delta d_{\rm L}/d_{\rm L},\Delta \Omega) =(0.1,~5~\mathrm{deg}^2)$ at $z\leq 0.1$ (left panel) and $(0.01,~1~\mathrm{deg}^2)$  at $z\leq 1$ (right panel). By comparing figure~\ref{fig:f2} and figure~\ref{fig:f1}, we can see that the constraint on $h$ becomes not only much less tight but also biased away from the input value, if the measurement errors of $d_{\rm L}$ and/or $\Delta \Omega$ are large.

Figure~\ref{fig:f3} shows the dependence of the precision of  $h$ and $\Omega_{\rm m}$ constraints on the luminosity distance and localization measurement errors of the GW events (totally $50$). As signified with blue markers, it is clear that the larger the luminosity distance errors (left panel) or the larger the localization errors (right panel), the poorer the constraint on $h$, and the constrained value of $h$ is even significantly biased away from the input value when $\delta d_{\rm L} /d_{\rm L}\gtrsim 0.1$ and $\Delta \Omega \gtrsim 1\,\mathrm{deg}^2$. 

To investigate the source for the systematic bias revealed in figures~\ref{fig:f2} and \ref{fig:f3}, we check the likelihood for a single GW event
\begin{equation}
p(\boldsymbol{D}_i|H,I) \propto \sum_{j=1}^m \exp\left[-\frac{\left(d_{\rm L}(z_j, H) - d_{L_i}\right)^2}{2\sigma^2_{d_{L_i}}}\right].
\label{eq:pdi}
\end{equation}

\begin{figure} 
\centering
\includegraphics[width=0.45\textwidth]{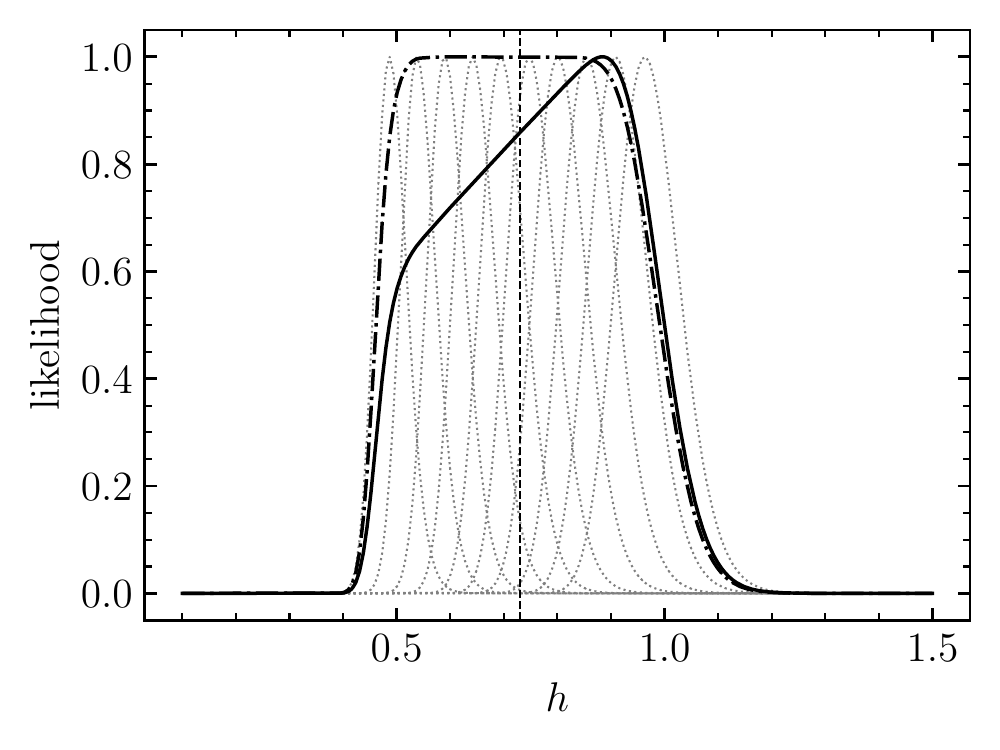}
\caption{Likelihood of individual GW events with different $h$ but fixed $\Omega_{\rm m}$ with uniform distribution of galaxies in the redshift error range. The dotted lines show the contributions from each galaxy in the redshift error range by fixing $\Omega_{\rm m}=0.25$, and the solid line marks the summation of the contributions from all these galaxies with peak rescaled to $1$. The vertical dashed line indicates the input value of the Hubble parameter. The dash-dotted line is the likelihood with consideration of the bias correction (see section~\ref{sec:result-bias}).
}
\label{fig:f4}
\end{figure}

We arbitrarily set some galaxies uniformly distributed in the redshift error range $[z^-, z^+]$ of a GW event and assume that each one has an equal probability to be the GW host. Then we calculate the likelihood according to the above equation for different cosmology parameters. Figure~\ref{fig:f4} displays the results for such a likelihood for cosmological models with various $h$ but fixed $\Omega_{\rm m}$ ($0.25$). In this figure, each gray line indicates the probability distribution of parameter $h$ given by a possible GW host galaxy with an assigned redshift in the redshift error range $z_j$, and the solid line represents the summation of the likelihood of all individual possible host galaxies  (signifying as the total likelihood for single event) with the peak rescaled to $1$ (see equation~\ref{eq:pdi}). Apparently, the total likelihood peaks at a value of $h$ larger than the input one and the likelihood distribution is not symmetric. The main reason is that the dependence of $d_{\rm L}$ on cosmological parameters ($h$, $\Omega_{\rm m})$ is nonlinear and the likelihood for individual host candidates is wider at large $h$ than that at small $h$, which causes the obtained constraints on the values of $h$ and $\Omega_{\rm m}$ to deviate from the input ones.

\begin{figure}
\centering
\includegraphics[width=0.5\textwidth]{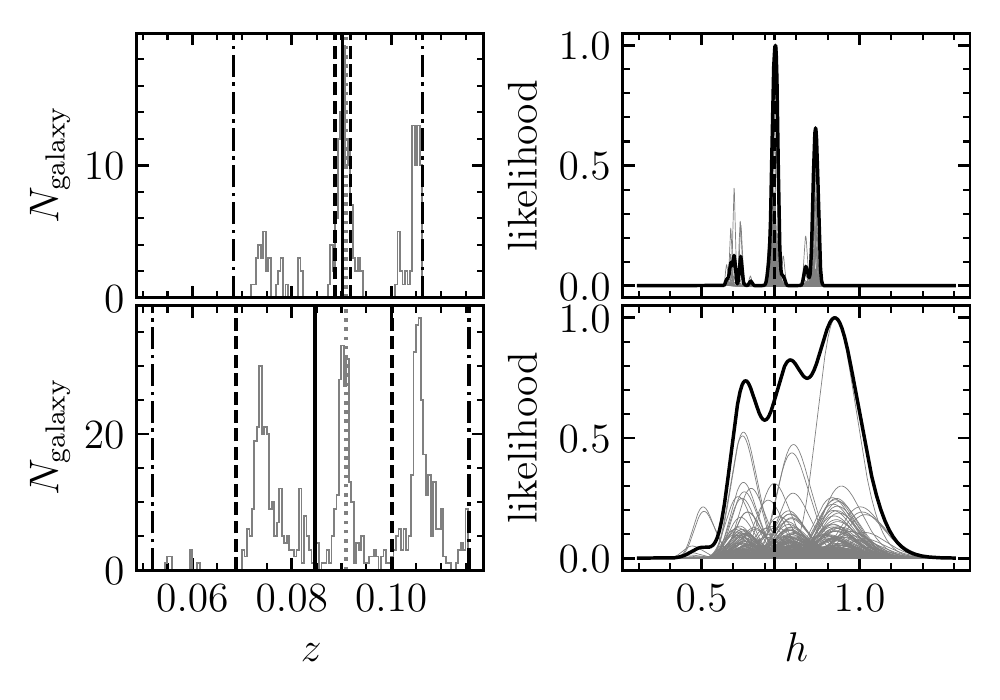}
\caption{
Galaxy distribution in the error volume/redshift error range of an example GW event (left panels) and the corresponding likelihood (right panels; $\Omega_{\rm m}=0.25$). Top and bottom panels show the two cases with settings of $(\delta d_{\rm L}/d_{\rm L}, \Delta \Omega) =(0.01,~1~\mathrm{deg}^2)$ and $(0.1,~5~\mathrm{deg}^2)$, respectively. The vertical dotted line indicates the true host redshift. The vertical solid line signifies the redshift converted from the best-fit value of luminosity distance $d_{\rm L}$ given by the GW observations and the two dashed lines indicate the redshift ranges corresponding to the $3\sigma$ range of $d_{\rm L}$, assuming the input cosmological model. The dash-dotted vertical line indicates the redshift range after accounting for the uncertainty of cosmology parameters and redshift uncertainty described in Section~\ref{sec:mockdata}. The gray lines show the likelihood contributed by each galaxy (weighted by mass) in the error volume, and the black line is the summation of all galaxies' contributions. The lines for each galaxy are scaled by their maximum value. The peak value of total likelihood is also scaled to $1$.
}
\label{fig:f5}
\end{figure}

In reality, galaxies in the redshift error range of a GW event may be clustered rather than uniformly distributed as assumed above. Figure~\ref{fig:f5} features the distribution of galaxies in the error redshift range of a typical GW event, which is clearly showing some cluster structures. One may also note that the true host may be not necessarily located in one of the clusters and at the center of redshift error range. When the luminosity distance error is large and thus error volume is large (bottom panels), many galaxies are potential hosts. The likelihoods of these galaxies are wide and they overlap with each other, which leads to a flat total likelihood over a large $h$ range with the peak biased to large $h$. When the luminosity distance error is small and thus the error volume is small (top panels), the number of galaxies that can be the potential host is limited. The likelihoods of these galaxies are narrow and the total likelihood of a GW event has several sharp peaks. By combining the total likelihood from different GW events, the contributions from the peaks without a ``true'' host cancel out and the one that is the ``true'' host is singled out, and thus a strong constraint on the cosmological parameters, including $h$ and $\Omega_{\rm m}$, can be obtained. 

We emphasize here that, in this dark siren method, the total likelihood for a single event is obtained by the summation of the likelihood of $h$ from individual galaxies in the error volume of each GW event. It is affected by the asymmetric nature and change in the width of the likelihood from individual galaxies, which is different from the case of standard sirens, where the corresponding effect is little because the estimation is obtained only by multiplying (rather than summing) the likelihoods contributed from individual host galaxies. A similar bias effect has been reported recently in \citet{Laghi2021} and \citet{Muttoni2021}. They found that low S/N events, with larger measurement errors, are not only less informative for inference of the cosmological parameters, but also tend to produce biased results. A non-Gaussian posterior can be helpful to mitigate this bias effect. A more realistic posterior may be considered for real GW events when using them as dark sirens though it is computationally expensive. In the present work, the likelihood is either assumed to be Gaussian or derived with the Fisher information matrix (FIM) approximation. The FIM approximation is only valid when S/N is high in which case it also assumes asymptotic Gaussian distributions for the errors of those estimated parameters.
 
\subsection{Correction of the bias}
\label{sec:result-bias}

In order to reduce the systematic bias, here we introduce a factor $\alpha(H)$ to re-balance the likelihood

\begin{eqnarray}
\alpha(H) \equiv \int_{z^{\min}}^{z^{\max}} dz ~ p_{\rm gal}(z|H) \delta(d_{\rm L_i} - \bar{d}_{\rm L}(z, H))\\\nonumber
= \frac{p_{\rm gal}(z(\bar{d}_{\rm L}, H)|H)}{\frac{d d_{\rm L}}{d z}|_{z=z(\bar{d}_{\rm L}, H)}},
\label{eq:alpha}
\end{eqnarray}
where $p_{\rm gal}(z|H)$ is the distribution of galaxies located in the error volume, and $\frac{d d_{\rm L}}{d z}= \frac{c(1+z)}{H_0E(z)} + \int_0^z \frac{c dz'}{H_0E(z')}$.

The likelihood for each individual GW event now can be written as
\begin{equation}
p(\boldsymbol{D}_i|H,I) \propto \frac{1}{\alpha(H)}\sum_{j=1}^m \exp\left[-\frac{\left(d_{\rm L}(z_j, H) - d_{L_i}\right)^2}{2\sigma^2_{d_{L_i}}}\right].
   \label{eq:pdi_debias}
\end{equation}

If we assume that galaxies are uniformly distributed in the redshift error range, then $\alpha(H) \propto 1/\frac{d d_{\rm L}}{d z}$, which is the case we considered in figure~\ref{fig:f4}. By including such a correction factor, the systematic bias in the likelihood can be removed as demonstrated by the dash-dotted line in figure \ref{fig:f4}.

For more realistic cases, we can still assume that galaxies are uniformly distributed in volume as long as the redshift range considered is not large. Therefore,
\begin{equation}
p_{\rm gal}(z|H) = \frac{1}{V_{\rm i}(H)} \frac{dV_{\rm c}}{dz} = \frac{4\pi}{V_{\rm i}(H)} \frac{\chi^2(z, H)}{H_0 E(z)}
\end{equation}
where $\chi(z, H)=\int_0^z \frac{c dz'}{H_0 E(z')}$ is the comoving distance and $V_{\rm i}(H)=\frac{4\pi}{3}[\chi^3(z^{\max}, H) - \chi^3(z^{\min}, H)]$ is volume between the redshift range of the error volume $[z^{\min}, z^{\max}]$. In principle, a more accurate estimation of $p_{\rm gal}(z|H)$ can be obtained by combining real galaxy survey observations rather than the simple assumption of uniform distribution of galaxies in volume.

With the above simple assumption, we re-run the previous cases by including the bias correction factor $\alpha(H)$, and also depict the results in figures~\ref{fig:f1}, \ref{fig:f2} and \ref{fig:f3}. Apparently, the constraints are only slightly improved by including the correction factor for those cases with high precision observations. However, the constraints can be significantly improved for those cases with low precision observations, and the biases can be removed for cases with modest measurement error.
While we also notice that with larger observational error (e.g. $(\delta d_{\rm L}/d_{\rm L}, \Delta \Omega) = (0.1,~1~\mathrm{deg}^2)$ for the high redshift case) the results are still biased, especially for $\Omega_m$, which need more investigation.

\begin{figure}
\centering
\includegraphics[width=0.5\textwidth]{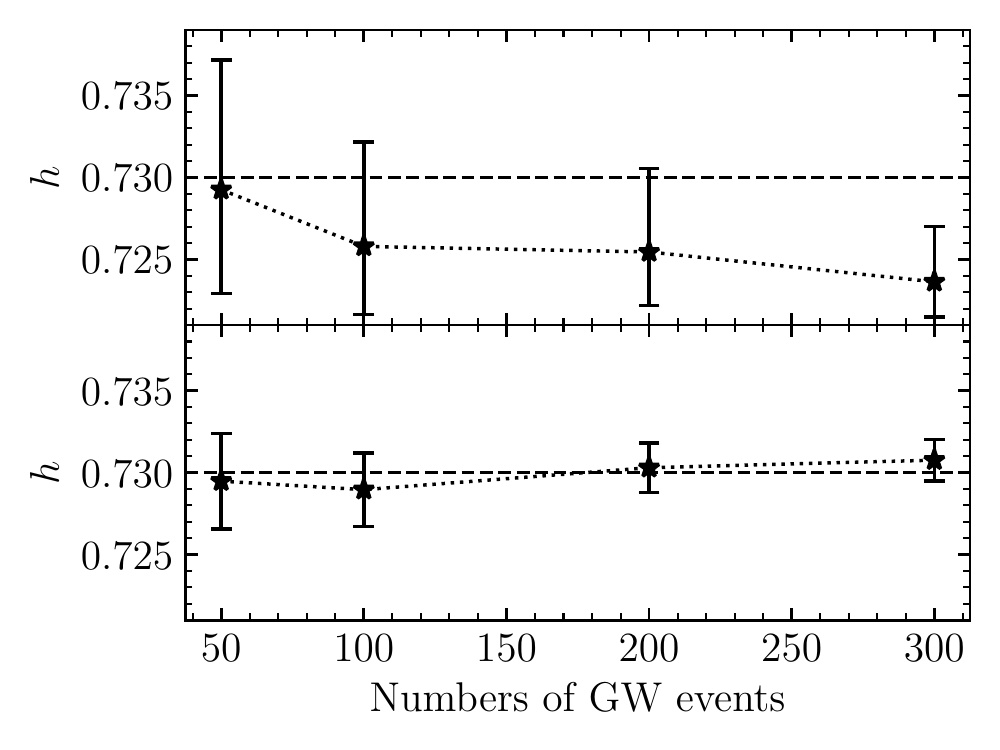}
\caption{
Constraints on $h$ versus the number of GW events for mock events at $z\leq 0.1$ with $(\delta d_{\rm L}/d_{\rm L},\Delta\Omega)=(0.01,~1~\mathrm{deg}^2)$. Top and bottom panels display the results with and without consideration of the peculiar velocity induced uncertainty in the redshift measurements of galaxies in the mocked catalog respectively.
}
\label{fig:f6}
\end{figure}

\begin{figure}
\centering
\includegraphics[width=0.5\textwidth]{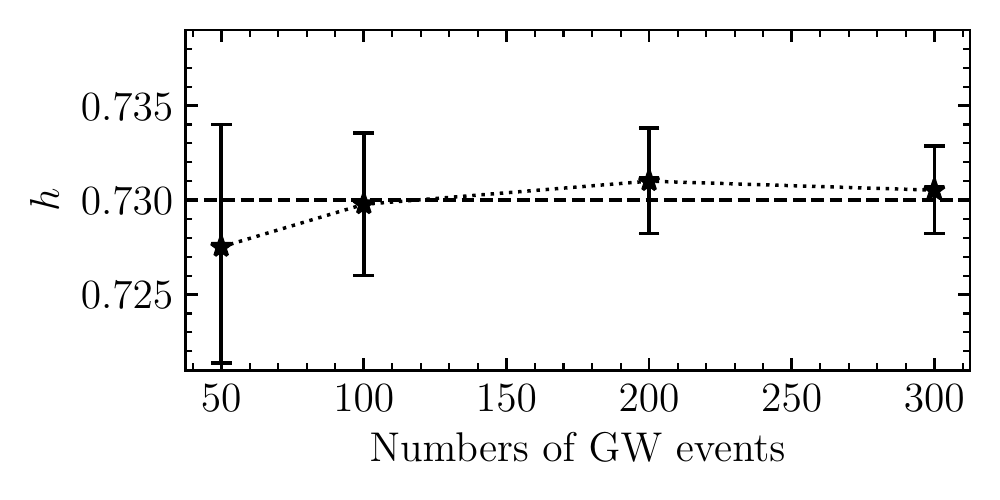}
\caption{
Constraints on $h$ versus the number of GW events for mock events at $z\leq 1$ with $(\delta d_{\rm L}/d_{\rm L},\Delta\Omega)=(0.01,~0.1~\mathrm{deg}^2)$.
}
\label{fig:f7}
\end{figure}

\subsection{Effect of event numbers and peculiar velocity}

One may expect that the constraints from dark sirens, with given measurement errors of the luminosity distance and localization, can be significantly improved if the number of available dark sirens increases to a much larger value. In order to check this, we further investigate the variation of constraints on the cosmological parameters with increasing number of BBHs that are adopted for the inference. 

Figure~\ref{fig:f6} displays the constraint on $h$ obtained from $50$, $100$, $200$, and $300$ mock GW events at $z<0.1$ with $(\delta d_{\rm L} /d_{\rm L},\Delta \Omega)=(0.01,~1~\mathrm{deg}^2)$ with (top panel) and without (bottom panel) consideration of the peculiar velocity induced error to galaxy redshift. It is clear that the constraint on $h$ becomes tighter with increasing number of adopted GW events and the statistical error of the constraint shrinks as expected (bottom panel), when both the luminosity distance and localization errors are sufficiently small. However, the best-fit value of $h$ may be significantly biased away from the input value with increasing number of GW events adopted in the analysis when the peculiar velocity induced redshift error is accounted for, though the statistical error indeed becomes smaller (top panel). If ignoring the peculiar velocities of host candidates, i.e., the redshift error range for a GW event is solely determined by $\delta d_{\rm L}/d_{\rm L}$, then the constraint on $h$ is improved with an error shrinking as expected ($\propto N^{1/2}$). This suggests that the constraint by BBHs at $z<0.1$ on $h$ may still be able to reach a precision substantially smaller than $1\%$ if the peculiar velocities of host candidates can be corrected on a galaxy-by-galaxy basis using the linear-multi-attractor model similar to that in \cite{Freedman2019}.

Figure~\ref{fig:f7} shows constraints on $h$ obtained from $50$, $100$, $200$ and $300$ mock GW events at $z<1$ with $(\delta d_{\rm L} /d_{\rm L},\Delta \Omega)=(0.01,~0.1~\mathrm{deg}^2)$. It is clear that $h$ is constrained well and the precision of the constraint improves with increasing number of GW events adopted. Since the BBHs in these samples span a large redshift range, the effects due to peculiar velocities are negligible. These suggest that only those GW events at high redshift with high precision $d_{\rm L}$ and $\Delta \Omega$ measurements, e.g., $\delta d_{\rm L} /d_{\rm L} \lesssim 0.01$ and $\Delta \Omega \lesssim 0.1~\mathrm{deg}^2$, can be used as dark sirens to tightly constrain the Hubble constant (and other cosmological parameters) with a precision $<1\%$.

\subsection{Effect of the host galaxy distribution prior}
\label{sec:discussion}

\begin{table*}
\centering
\caption{Effects on the constraints to $h$ and $\Omega_{\rm m}$ by choosing different settings for the properties of GW host galaxies.
}
\resizebox{\textwidth}{!}{
\renewcommand{\arraystretch}{1.5}
\begin{tabular}{c|cccccccc}\hline \hline
\multirow{3}{*}{True $p_j$} & \multicolumn{8}{c}{Prior $p_j$} \\ \cline{2-9}
& \multicolumn{2}{c}{mass-weight} &    &  \multicolumn{2}{c}{equal-weight} & & \multicolumn{2}{c}{log-normal-weight} \\ \cline{2-3} \cline{5-6} \cline{8-9}
& $h$ & $\Omega_{\rm m}$ & & $h$ & $\Omega_{\rm m}$ & & $h$ & $\Omega_{\rm m}$ \\ \hline
mass-weight & $0.7307_{-0.0065}^{+0.0077}$ & $0.2829_{-0.1578}^{+0.1545}$ & & $0.7302_{-0.0066}^{+0.0077}$ & $0.2991_{-0.1721}^{+0.1373}$ & & $0.7296_{-0.0064}^{+0.0078}$ & $0.2973_{-0.1699}^{+0.1377}$ \\  \hline
log-normal-weight & $0.7301_{-0.0094}^{+0.0083}$ & $0.2038_{-0.1198}^{+0.1481}$ & & $0.7291_{-0.0093}^{+0.0091}$ & $0.2293_{-0.1333}^{+0.1442}$ & & $0.7309_{-0.0094}^{+0.0084}$ & $0.2071_{-0.1222}^{+0.1439}$ \\ \hline \hline
\end{tabular}}
\label{tab:t1}
\end{table*}

In our analysis in section~\ref{sec:result}, we simply assume the probability of a galaxy in the redshift error range is proportional to the galaxy mass. However, the probability of a galaxy to be the host may deviate from this simple assumption as not only the formation of BBHs may depend on other properties of galaxies, e.g., metallicity and morphology \citep{Cao2018, Artale2019}, but also different channels for the origin of BBHs would lead to a significantly different distribution of BBH host galaxies. This probability is not known currently, though it may be tightly constrained with the accumulation of BBH detection in the future. It may also be inferred by using models for the formation and evolution of BBHs across the cosmic time but it depends on many detailed physics that are currently not well understood. Here we check whether the constraints on cosmological parameters are dependent on the choice of the prior for this distribution. To do so, we arbitrarily set three different types of weight $p_j$ for galaxies in the redshift error range of a GW event: 1) mass-weight, i.e., the probability for a galaxy in the redshift error range of a GW event is proportional to its stellar mass (the default model if not otherwise stated); 2) equal-weight, i.e., all galaxies within the redshift error range have equal weight, and the probability for each of those galaxies to be the host is independent of any galaxy property; 3) log-normal-weight, i.e., the probability distribution of a GW event in a galaxy with mass in the range $M_\ast \rightarrow M_\ast+dM_\ast$ is assumed to be $dN/d\log M_\ast \propto \exp\left[ -\frac{1}{2} \left(\frac{\log (M_\ast/3\times10^{10}M_\odot)}{\log2}\right)^2 \right]$  (c.f., the host galaxy distributions obtained in \citet{Cao2018}). If adopting the last one, for each GW event, we rank those galaxies in the redshift error range with mass from small to large as $j=1, \cdots, n$, and assign probability for the $j$-th galaxy as $p_j(M_{\ast,j}) = \int_{(M_{\ast,j-1}+ M_{\ast,j})/2}^{(M_{\ast,j} +M_{\ast,j+1})/2} dM_\ast dN/dM_\ast$, and the lower limit of the integration is set as $0$ when $j=1$, while the upper limit is set as $\infty$ when $j=n$.

We adopt the above settings for the weight $p(M_\ast)$ to generate mock catalogs of $50$ GW events and their host galaxies at redshift $z<0.1$. Similar to the previous analysis, we assume $\delta d_{\rm L}/d_{\rm L}=0.01$ and $\Delta \Omega=1\,\mathrm{deg}^2$. For each mock catalog with a given setting for $p_j$, we use Bayes' inference to constrain the cosmological parameters by assuming either the ``true'' $p(M_\ast)$ or the other two types. Our results are listed in Table~\ref{tab:t1}, which suggest that the setting on the prior for the dependence of $p_j$ (probability of a galaxy to be the host) on galaxy properties has little effect on the constraints of cosmological parameters ($h$ and $\Omega_{\rm m}$). The reason might be that (1) the redshift error range for each GW event is large and different types of galaxies can always be selected with representative numbers, and (2) the peculiar velocities of galaxies may lead to mixing of the observed redshift distributions for different types of galaxies and thus there will always be different types of host candidates that have redshift close to the ``true'' host. Therefore, we conclude that the Bayes' inference of cosmological parameters is robust even if no information about the properties of host galaxies is provided. This supports the proposal of using galaxy groups as a surrogate for individual host galaxies to constrain cosmological parameters in \citet{Yu2020}.

\section{Expected cosmological constraints from Deci-hertz Observatory}
\label{sec:DO}

We have demonstrated above that dark sirens with sufficiently small luminosity distance and localization errors can put strong constraints on cosmological parameters, such as $h$ and $\Omega_{\rm m}$. Future GW observations may provide a large number of dark sirens but not all of them can have precise measurements on their luminosity distances and localizations. For example, most BBHs that will be detected by the third generation of ground-based GW detectors are expected to be localized at $\Delta \Omega \gtrsim 0.1~\mathrm{deg}^2$ and have luminosity distance measurement errors of $\delta d_{\rm L}/d_{\rm L} \gtrsim 0.01$ and the number of BBHs that have sufficiently small $d_{\rm L}$ and $\Delta \Omega$ errors is limited \citep{Taylor2012, Vitale2017a, Nair2018, Zhao2018}. 
However, DO and DECIGO are anticipated to detect a large number of BBHs with extremely high S/N and sufficiently small errors in $d_{\rm L}$ and localization measurements. Therefore, in this section, we investigate whether DO BBHs can be used as dark sirens to put strong constraints on $h$ and $\Omega_{\rm m}$ by using a mock BBH sample of DO observations. We first generate a mock sample of BBHs according to the merger rate density and its evolution $R(z,M_{\rm c})=R(z)P(M_{\rm c})$, here $M_{\rm c} = (m_1 m_2)^{3/5}/(m_1+m_2)^{1/5}$ is the chirp mass of BBH, and $P(M_{\rm c})$ is obtained by using the primary mass ($m_1$) and mass ratio ($q=m_2/m_1$) distributions, given in \citet{ligoO3a} for the power law plus a peak model. With this $R(z, M_{\rm c})$, the number distribution for BBHs within the period range from $P$ to $P+dP$ at redshift $z$ is then given by \citep{ZhaoLu2021}
\begin{equation}
\frac{dN}{dP} \simeq
\int \frac{5 R(z,M_{\rm c})c^5  P^{5/3}}{384 \cdot 2^{2/3}\pi^{8/3}(G M_{\rm c})^{5/3}} \frac{dV}{dz}  d M_{\rm c}.
\label{eq:num-circ}
\end{equation}
Then, we can generate a mock sample of BBHs by the Monte Carlo method and for each mock BBH in the sample, its mass, mass ratio and GW frequency at the starting time ($f_{\rm i}$) of DO observations can be obtained. Considering that DO is extremely sensitive for detecting BBHs, we first consider a half year of observations and generate a mock BBH sample that can be detected by DO with an S/N ($\varrho$) threshold of $8$.

The S/N of each BBH is given by \citep{Maggiore2008}
\begin{align}
\label{eq:snr}
\varrho^2 = 2\left(h\mid h\right) = 2 \int_{f_{i}}^{f_{f}} \frac{4\tilde{h}^{*}(f) \tilde{h}(f)}{S_{n}(f)} d f
= 2\int_{f_i}^{f_f} \frac{h_{\rm c}^2(f)}{f S_{\rm n}(f)}d\ln f. \nonumber \\
\end{align}
Here the factor of $2$ accounts for the two Michelson interferometers of DO, $S_{\rm n}(f)$ is the non-sky-averaged single-detector noise power density of DO-Conservative taken from \citet{ArcaSedda2020} and $h_{\rm c}(f)=2f|\tilde{h}(f)|$ is the characteristic strain where $\tilde{h}(f)$ is the Fourier transform of GW  signal $h(t)$. Adopting the Newtonian approximation for the inspiral stage of BBHs \citep{Maggiore2008}, we have
\begin{equation}
\tilde{h}(f)=\left(\frac{5}{24}\right)^{1 / 2} \frac{1}{\pi^{2 / 3}} \frac{c}{d_{\rm L}}\left(\frac{G \mc}{c^{3}}\right)^{5 / 6} f^{-7 / 6} Q e^{i \Psi(f)}.
\label{eq:h_f}
\end{equation}
Here $\mc=(1+z)M_{\rm c}$ is the redshifted chirp mass, 
$f=f_r/(1+z)$ is the observed frequency, $\Psi(f)$ is the strain phase, $Q$ is a quantity related to the detector's pattern function $F^+$, $F^\times$, $\iota$ is the angle between the angular momentum of the source and the vector pointing from the detector to the source in the detector's frame, and
\begin{align}
Q(\theta, \phi, \psi ; \iota) = \sqrt{F^2_{+} \left(\frac{1+\cos ^{2} \iota}{2}\right)^2+ F^2_{\times} \cos^2 \iota}, 
\end{align}
with
\begin{align}
F_{+}(\theta, \phi, \psi) = ~& \frac{1}{2}\left(1+\cos ^{2} \theta\right) \cos 2 \phi \cos 2 \psi \nonumber \\
& -\cos \theta \sin 2 \phi \sin 2 \psi, \\
F_{\times}(\theta, \phi, \psi) = ~& \frac{1}{2}\left(1+\cos ^{2} \theta\right) \cos 2 \phi \sin 2 \psi \nonumber \\
& +\cos \theta \sin 2 \phi \cos 2 \psi.
\end{align}
Note here $\theta$, $\phi$, $\psi$ and $\iota$ are the polar angle, azimuthal angle, polarization angle and inclination angle of the source in the detector frame respectively.
After averaging over all possible directions and inclinations, we have
\begin{equation}
\left\langle|Q(\theta, \phi, \psi ; \iota)|^{2}\right\rangle^{1 / 2}=\frac{2}{5}.
\label{eq:Q}
\end{equation}

Combining Equations~\eqref{eq:snr}, \eqref{eq:h_f} and \eqref{eq:Q}, the averaged S/N is
\begin{align}
\varrho =
\frac{2}{\sqrt{15} \pi^{\frac{2}{3}}}
\frac{c}{d_{\rm L}}\left(\frac{G\mc}{c^{3}}\right)^{\frac{5}{6}} 
\left[\int_{f_{\rm i}}^{f_{\rm f}} df \frac{f^{-\frac{7}{3}}}{S_{n}(f)}\right]^{\frac{1}{2}},
\end{align}
where $f_{\rm i}$, assigned according to the orbital period distribution in equation~\eqref{eq:num-circ}, is the initial frequency of the mock BBH at the beginning of DO observation, $f_{\rm f}= \min(10\,{\rm Hz}, f_{\rm ISCO})$ and $f_{\rm ISCO}= 1/(6^{3/2}\pi M/M_\odot)$\,Hz.

For DO detectors, we take the non-sky-averaged waveform in the parameter estimation. Their orbital motions during the observation are functions of time $t$. We change the detector frame parameters $(\theta,\phi,\psi)$ to ecliptic coordinates. With the source position denoted by $(\theta_{\rm S},\phi_{\rm S})$ and the orbital angular momentum direction of the mock BBH denoted by $(\theta_{\rm L},\phi_{\rm L})$, we can then use $(\theta_{\rm S},\phi_{\rm S},\theta_{\rm L},\phi_{\rm L})$ to replace $(\theta,\phi,\psi,\iota)$ (see equations (10)-(19) in \citet{Liu2020}  or equations (3.16-3.22) in \citet{Cutler1998} ). We randomly assign these four values for each object in the mock sample. In the frame transformation, there are two important parameters: (1) the azimuthal angle of the detector around the Sun $\Phi(t)=\Phi_0+\frac{2\pi t(f)}{T}$, where $T$ is the orbital period of the detector that is equal to one year; and (2)  the initial orientation of the detector arms $\alpha_0$.

The expected uncertainties in the measurements of the BBH parameters $\boldsymbol{\Xi} =\left\{d_{\rm L}, \mc, \eta, t_{\rm c}, \phi_{\mathrm{c}}, \theta_{\rm S}, \phi_{\rm S}\right\}$ may be estimated by using the FIM method, where $\eta=m_1m_2/(m_1+m_2)^2$. The FIM can be obtained as
\begin{align}
\Gamma_{a b} & =  \left(\left.\frac{\partial h}{\partial \boldsymbol{\Xi}^{a}} \right| \frac{\partial h}{\partial \boldsymbol{\Xi}^{b}}\right) \nonumber \\
& =  2\sum_{j=1}^{2} \int_{f_{\rm i}}^{f_{\rm f}} \frac{\frac{\partial \tilde{h}_j^{*}(f)}{\partial \boldsymbol{\Xi}^{a}}\frac{\partial \tilde{h}_j(f)}{\partial \boldsymbol{\Xi}^{b}}+\frac{\partial \tilde{h}_j^{*}(f)}{\partial \boldsymbol{\Xi}^{b}}\frac{\partial \tilde{h}_j(f)}{\partial \boldsymbol{\Xi}^{a}}}{S_{n}(f)} df,
\end{align}
where $j$ refers to detector ``1" (with $\alpha_0=0$, $\Phi_0=0$) or ``2" (with $\alpha_0=\pi/4$,  $\Phi_0=0$). Given $\Gamma^{ab}$, then we have
\begin{equation}
\left\langle\delta \Xi^{a} \delta \Xi^{b}\right\rangle=\left(\Gamma^{-1}\right)^{a b},
\end{equation}
and thus we can estimate the uncertainties in the measurements of $\boldsymbol{\Xi}$ as
\begin{equation}
\Delta \boldsymbol{\Xi}^{a}=\sqrt{\left(\Gamma^{-1}\right)^{a a}}.
\end{equation}
In particular, the angular resolution $\Delta\Omega$ is defined as
\begin{equation}
\Delta\Omega=2\pi|\sin\theta_{\rm S}|\sqrt{(\Delta\theta_{\rm S}\Delta\phi_{\rm S})^2-\left\langle\Delta\theta_{\rm S}\Delta\phi_{\rm S}\right\rangle^2}
\end{equation}

For both DO detectors, the GW strain signal can be described by equation~\eqref{eq:h_f}. The GW strain phase evolution includes the polarization modulation and Doppler modulation
\begin{align}
\Psi(f) =~& 2 \pi f t_{\rm c}-\phi_{\rm c}-\pi/4-\phi_{\rm p}-\phi_{\rm D}+\frac{3}{4}[8 \pi \mc f]^{-5/3} \nonumber \\
&  \times\left[1+\frac{20}{9}\left(\frac{743}{336}+\frac{11\eta}{4}\right) x-16\pi x^{3/2}\right],
\end{align}
where $x(f) \equiv[\pi M(1+z) f]^{2 / 3}$.
The polarization modulation can be written as
\begin{equation}
\phi_{\rm p}(t(f))=\arctan \frac{-2 \cos \imath F_{\times}(t(f))}{\left(1+\cos ^{2} l\right) F_{+}(t(f))}.
\end{equation}
The motion of DO detectors around the helio-center causes Doppler modulation of the GW phase as
\begin{equation}
\phi_{\rm D}(t(f))=2 \pi f R \sin \theta_{\rm S} \cos \left(\Phi(t(f))-\phi_{\rm S}\right),
\end{equation}
where $R=1$ AU is the orbital radius of the detector and
\begin{align}
t(f)=~& t_{\rm c}-5(8 \pi f)^{-8 / 3}\mc^{-5 / 3} \\
& \times\left[1+\frac{4}{3}\left(\frac{743}{336}+\frac{11 \eta}{4}\right) x-\frac{32 \pi}{5} x^{3 / 2}\right].
\end{align}

\begin{figure}
\centering
\includegraphics[width=0.5\textwidth]{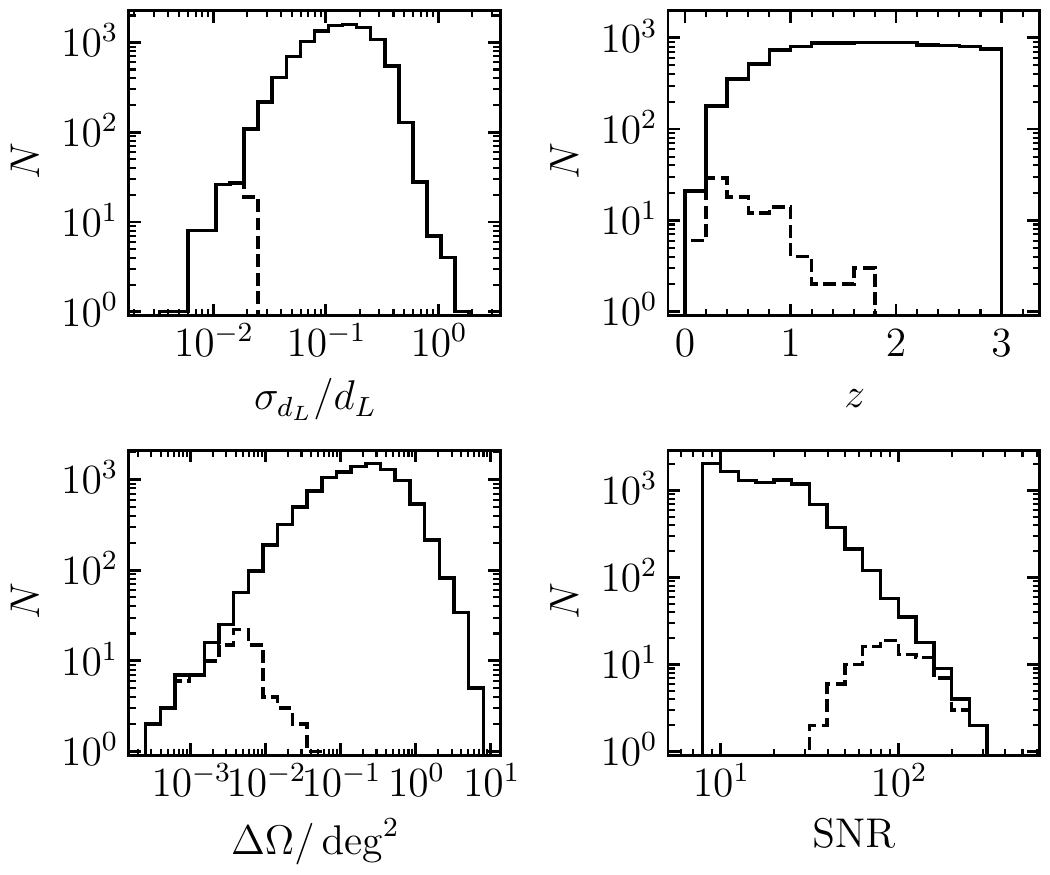}
\caption{
Distributions of mock BBH mergers expected to be detected by DO. Solid and dashed histograms display the distribution of the whole mock sample of BBHs ``detected'' by DO with S/N$>8$ in a half year observation period (totally $10269$) and that of a subsample with $\sigma_{d_{\rm L}}/ d_{\rm L}< 0.02$ (totally $90$) adopted for cosmological parameter inference, respectively.
}
\label{fig:f8}
\end{figure}

With the above formalism, we can estimate S/N values for all the mock BBHs that can be detected ($\varrho>8$) by DO within a half year observation and the precision of parameter measurements for each ``detectable'' mock BBH. For all those ``detectable'' BBHs, Figure~\ref{fig:f8} plots the distributions of their S/N (bottom-right panel), redshift (top-right panel), measurement errors of luminosity distance (top-left panel) and localization (bottom-left panel). Apparently more than hundreds of mock BBHs can be detected with $\varrho \gtrsim 30$, which basically have  accurate determinations of their luminosity distances and locations. The dashed histograms in this figure show the distributions of those mock BBHs (totally $90$ and typically at $z\lesssim 2$) that have $\sigma_{d_{\rm L}}<0.02$ and $\Delta \Omega \lesssim 0.1~\mathrm{deg}^2$. This subsample may be used to do cosmological inference as we have demonstrated in section~\ref{sec:result} that the mock BBHs with precise luminosity distance measurements and localization can be used to obtain tight constraints on the Hubble constant and other cosmological parameters. 

\begin{figure}[!hbtp]
\centering
\includegraphics[width=0.49\textwidth]{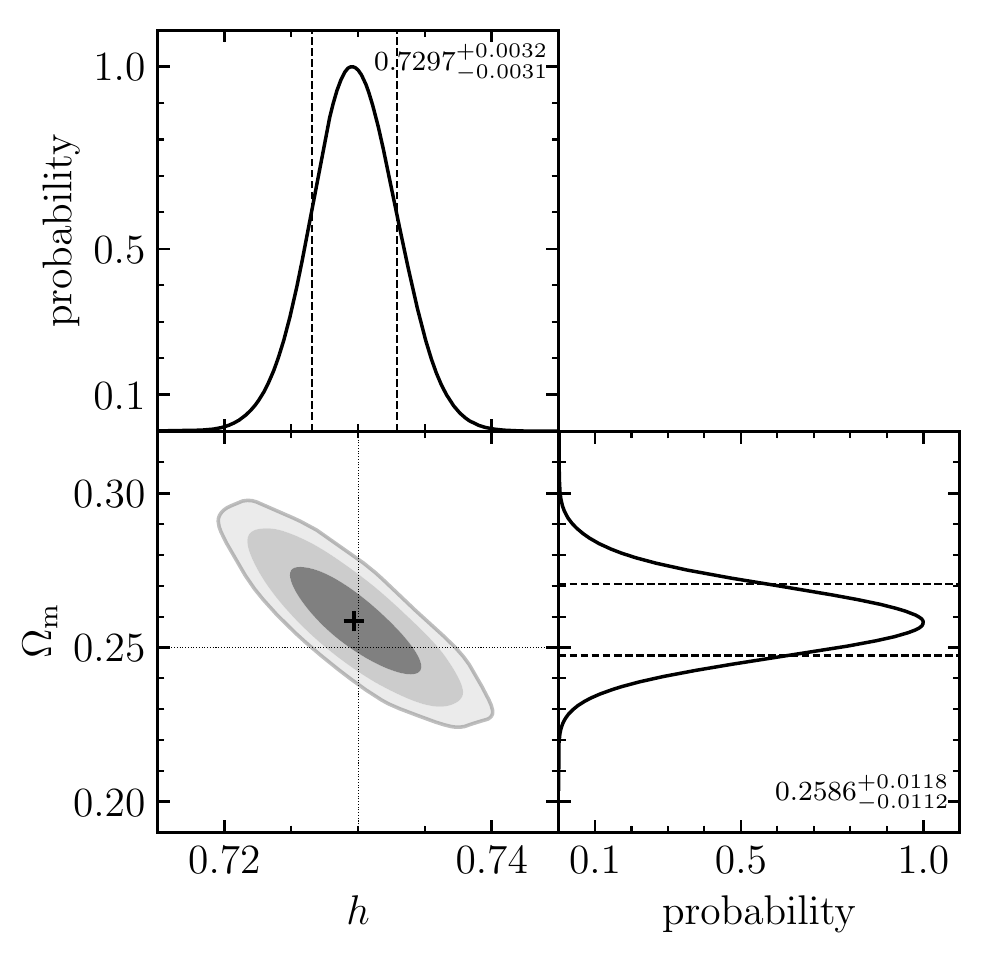}
\caption{
Posterior probability distributions of $h$ and $\Omega_m$ obtained from the mock BBH mergers ``detected'' by DO with luminosity distance uncertainties of $\delta {d_{\rm L}}/d_{\rm L} < 0.02$.
}
\label{fig:f9}
\end{figure}

Figure~\ref{fig:f9} depicts the constraints on $h$ and $\Omega_{\rm m}$ obtained from those $90$ DO BBHs with $\sigma_{\rm d_{\rm L}}<0.02$. Apparently, the constraint on $h$ can be as tight as $0.44\%$. However, $\Omega_{\rm m}$ is less tightly constrained and the accuracy is only $\sim 4-5\%$. Note here we only consider the case of a half year observation of DO. The number of ``detected'' BBHs with $\sigma_{d_{\rm L}}<0.02$ or smaller increases with increasing observation time ($\propto T_{\rm obs}$), which means that the constraints  on $h$ can be improved as $\propto T_{\rm obs}^{-1/2}$. If $T_{\rm obs} =10$\,yr, the constraint on $h$ and $\Omega_{\rm m}$ may reach $\sim 0.1\%$ and $\sim 1\%$, respectively. 

Note here that we do not consider the bias correction introduced in section~\ref{sec:result-bias} as the DO sample we adopted above are all have precise measurements on the luminosity distances ($\delta d_{\rm L}/d_{\rm L} <0.02$) and localization areas ($\delta \Omega \lesssim 0.01~\mathrm{deg}^2$). If including sources with larger luminosity distance errors and localization errors, one may need to consider the bias correction.

For comparison, the accuracy of recent $h$ measurements by using the local distance ladder is about 2\% level \citep[e.g.,][]{Riess2019}. Surveys such as those with Euclid, the Square Kilometre Array (SKA), eBOSS and DESI are expected to reach sub-percent precision on Baryon Acoustic Oscillation (BAO) measurements of the Hubble parameter at redshift $0.1<z<3$ and give constraints on $h$ close to 1\% in the best cases \citep{Wang2017, Bengaly2020}. As for GW standard sirens, \citet{Chen2018} predict a 2\% measurement of $h$ with Advanced LIGO-Virgo within five years, and expected reach about 0.3\% for future ET with 10 years of observation \citep{Zhang2019}. In addition, with five-years of observation by future space-based GW detector Taiji, the constraint on $h$ is expected to be $\sim 1.3\%$ using massive BBH as standard sirens and $\sim 1.0\%$ with the network of LISA and Taiji \citep{Wang2021}. This suggests that, with the observation of DO, stellar mass BBHs as dark sirens, as an independent measurement of cosmological parameters, is quite promising.

\section{Conclusions}
\label{sec:conclusion}

GW events can be used as standard sirens to constrain cosmological parameters if their redshift can be measured by EM observations. Although a large fraction of GW events may not have EM counterparts and cannot have direct redshift measurements, e.g.,  BBH mergers, one may still use it as dark standard sirens to statistically constrain cosmological parameters provided that the redshifts of their host candidates can be obtained from deep galaxy surveys. In this paper, we investigate this dark siren method in detail by using mock GW events and galaxy catalogs. We find that the Hubble constant can be constrained well (on percentage level or better) by using a few tens or more of BBH mergers at redshift up to $1$ with accurate luminosity distance measurements (relative error $\lesssim 1\%$) and localization (sky coverage $\lesssim 0.1\mathrm{deg}^2$), though the constraint may be significantly biased if these measurements are less accurate and the bias cannot be removed by simply increasing the number of adopted GW events. 
We analyze this systematic bias in detail and introduce a correction method to remove it. We demonstrate that this correction method works well in most cases, though it may not work well when the measurement errors of GW luminosity distances and localization areas are too large. Our results suggest that GW events as standard dark sirens can be utilized to robustly constrain the cosmological parameters once their luminosity distances and localization areas are well determined. So, selection of sources with sufficiently high precision measurements of the luminosity distances and localization areas is important for obtaining robust constraints on the cosmological parameters.

Finally, we simulate future observations of BBH inspirals and mergers by DO, according to current constraints on BBH merger rate and distributions of BBH properties. We find that DO in a half year observation period may detect about one hundred BBHs with S/N $\varrho \gtrsim 30$, relative luminosity distance error $\lesssim 0.02$, and localization error $<0.01\mathrm{deg}^2$.  We predict that the Hubble constant can be constrained to the $\sim 0.1-1\%$ level using these DO BBHs as dark sirens, which is comparable with the accuracy of other methods in the near future. This dark siren method, independent of those methods that rely on the local distance ladder, BAO and CMB observations, would provide insight into the tension between the early- and late-Universe measurements of the Hubble constant. We also demonstrate that the constraint on the Hubble constant using this dark siren method is robust and does not depend on the choice of the prior for the properties of BBH host galaxies.

\normalem
\begin{acknowledgements}
This work is partly supported by the National Key Program for Science and Technology Research and Development (Grant Nos. 2020YFC2201400, 2020SKA0120102, and 2016YFA0400704), the National Natural Science Foundation of China (Grant No. 11690024), and the Strategic Priority Program of the Chinese Academy of Sciences (Grant XDB 23040100).
\end{acknowledgements}
  
\bibliographystyle{raa}

\end{document}